\documentclass[10pt,conference]{IEEEtran}
\usepackage[utf8]{inputenc}
\usepackage[T1]{fontenc}
\usepackage{graphicx,color,mathenv}
\usepackage[cmex10]{amsmath}
\usepackage{cite,amsthm,amssymb}
\usepackage{balance}
\usepackage[ruled,vlined]{algorithm2e}
\usepackage{subfigure,amsfonts}
\usepackage{float}
\usepackage{url}
\usepackage{xcolor}

\usepackage{bbm}
\usepackage{cancel}

\DeclareMathAlphabet\mathbfcal{OMS}{cmsy}{b}{n}

\begin{document}
\bstctlcite{IEEEtranBSTCTL}
\title{Joint PAPR and OOBE Reduction for AFDM \\ via Chirp Parameter Tuning}
%\vspace{-3em}

\author{\IEEEauthorblockN{
Vincent Savaux\IEEEauthorrefmark{1},
Hyeon Seok Rou\IEEEauthorrefmark{2}, 
Zeping Sui\IEEEauthorrefmark{3}, 
Giuseppe Thadeu Freitas de Abreu\IEEEauthorrefmark{2}, and 
Zilong Liu\IEEEauthorrefmark{3} \vspace{1ex}
}
\IEEEauthorblockA{\IEEEauthorrefmark{1}b$<>$com, 1219 Av. des Champs Blancs, 35510 Cesson-S\'{e}vign\'{e}, France}
\IEEEauthorblockA{\IEEEauthorrefmark{2}School of Computer Science and Engineering, Constructor University, Campus Ring 1, 28759 Bremen, Germany}
\IEEEauthorblockA{\IEEEauthorrefmark{3}School of Computer Science and Electronics Engineering, University of Essex, Colchester CO4 3SQ, U.K.}
\IEEEauthorblockA{
Email: vincent.savaux@b-com.com, \{hrou,gabreu\}@constructor.university, \{z.sui,zilong.liu\}@essex.ac.uk}
\vspace{-1.5em}
}

        % <-this % stops a space


\maketitle

\begin{abstract}
This paper addresses the joint reduction of the peak-to-average power ratio (PAPR) and out-of-band emissions (OOBE) in affine frequency division multiplexing (AFDM) systems by selecting the pre-chirp parameter $c_2$. While existing approaches typically optimize either PAPR or OOBE independently, the proposed method jointly considers both metrics. To this end, a weighted cost function combining PAPR and OOBE is introduced to evaluate the trade-off between the two objectives. A pre-chirp selection scheme, inspired by the selected mapping (SLM) technique, is then employed to identify the optimal $c_2$ value from a finite set of candidates, yielding a Pareto-optimal operating point within a discrete set. Simulation results demonstrate that the proposed approach simultaneously reduces both PAPR and OOBE compared with conventional AFDM. Moreover, its performance remains close to that of methods specifically optimized for a single objective, with only about a 1 dB degradation in PAPR reduction and a 2–3 dB degradation in OOBE suppression.    
\end{abstract}


%%%%%%%%%%%%%%%%%%%%%%%%%%%%%%%%%%%%%%%%%%%%%%%%%%%%%%%%%%%%%%%%%%%%%%%%%%%%%%%%%%%%%%%%%%%%%%%%%%%%%%%%%%%%%%%%%%%%%%%%%%%%%%%%%%%%%%%%%%%%%%%%%%%%%%%%%%%%%%%%%%%%%%%%%%%%%%%%%%%%%%%%%%%%%%%%%%%%%%%%%%%%%%%%%%%%%%%%%%%%%%%%%%%%%%%%%%%%%%%%%%%%%%%%%%%%%%%%%%%%

\vspace{1.5em}
\section{Introduction}
\label{sec:intro} 

The increasing demand for reliable wireless communications in high-mobility scenarios, such as high-speed railways, unmanned aerial vehicles (UAVs), and non-terrestrial networks (NTNs), has exposed the limitations of orthogonal frequency division multiplexing (OFDM) \cite{sui2025multi}. While OFDM remains the waveform of choice in current cellular standards, its sensitivity to Doppler spread severely limits its performance in rapidly time-varying environments \cite{R12508043}. Orthogonal time frequency space (OTFS) has recently attracted considerable attention by representing information in the delay-Doppler domain \cite{wei21}, thereby transforming doubly dispersive channels into a sparse representation and improving resilience against mobility \cite{10250854}. Nevertheless, the two-dimensional modulation and detection operations of OTFS may introduce additional implementation complexity. In this context, affine frequency division multiplexing (AFDM) has recently been proposed as a novel chirp-based multicarrier waveform that leverages the discrete affine Fourier transform (DAFT) to efficiently combat both delay and Doppler dispersions \cite{bemani21,bemani23} while maintaining a transceiver architecture of complexity comparable to OFDM \cite{savaux24ieeetcom}. Owing to its ability to exploit full delay-Doppler diversity and its compatibility with existing multi-carrier processing techniques, AFDM has emerged as a compelling waveform for future high-mobility wireless communication systems and networks \cite{11556327,AFDM_CSM25}.

However, like OFDM, AFDM suffers from a high peak-to-average power ratio (PAPR) and significant out-of-band emissions (OOBE). Several approaches have recently been proposed to mitigate these drawbacks. In \cite{ali25}, spreading matrices are applied prior to AFDM modulation to reduce the PAPR, although only limited performance gains are achieved. 
Alternatively, filter-bank multicarrier (FBMC)-inspired augmentations of AFDM have been investigated, leveraging the superior PAPR and OOBE characteristics of FBMC while retaining the chirp-based benefits of AFDM \cite{AFBM_TWC26}, but such methods incur additional overhead.
In \cite{yuan25,choi26}, a pre-chirp selection scheme, analogous to the selected mapping (SLM) technique used in OFDM, is proposed. The authors of \cite{reddy25} employ $\mu$-law companding for PAPR reduction and extend this work in \cite{reddy26} by combining pre-chirp selection with $\mu$-law companding in a hybrid framework. In \cite{lu25}, a precoding technique inspired by single-carrier frequency division multiplexing (SC-FDM) is proposed, providing an intrinsic reduction in PAPR. 

In this paper, we focus on the pre-chirp selection approach because of its low implementation complexity and its flexibility for integration with AFDM systems.
Specifically, most of the aforementioned approaches focus exclusively on PAPR reduction, whereas other physical-layer performance metrics may also be of interest. We extend the pre-chirp selection method to jointly minimize both PAPR and OOBE. In fact, since the pre-chirp is applied in the affine domain, it can be chosen to reduce the OOBE which is assessed in the frequency domain. To this end, we introduce a new cost function based on the weighted Euclidean norm of the normalized PAPR and OOBE, defined as the square root of their weighted quadratic sum. The weighting factor provides a flexible trade-off between the two objectives by assigning different priorities to each metric. The optimal pre-chirp parameter is then selected by minimizing this cost function. Simulation results demonstrate that, when PAPR and OOBE are equally weighted, the proposed method achieves simultaneous reductions in both metrics compared with conventional AFDM. Furthermore, its performance approaches that obtained when the pre-chirp selection is specifically optimized for either PAPR or OOBE alone. Finally, the results show that the performance further improves as the size of the candidate set for the pre-chirp parameter increases. 

% The rest of the paper is organized as follows: Section \ref{sec:system_model} presents the AFDM signal model. The proposed joint PAPR and OOBE minimization method is introduced in Section \ref{sec:perf_analysis}. The simulation results are provided in Section \ref{sec:simu}, and Section \ref{sec:conclusion} concludes this paper. 
%
The rest of the paper is organized as follows: Section \ref{sec:system_model} presents the AFDM signal model and the system model, Section \ref{sec:perf_analysis} introduces the proposed joint PAPR and OOBE minimization method, Section \ref{sec:simu} provides simulation results, and Section \ref{sec:conclusion} concludes the paper.

\textit{Notations}: Boldface lowercase $\mathbf{a}$ and uppercase $\mathbf{A}$ denote vectors and matrices, respectively, while normal font $a$ denotes scalar variables. $\mathbf{A}^{*}$, $\mathbf{A}^{T}$, and $\mathbf{A}^{H}$ denote the conjugate, transpose, and conjugate transpose (Hermitian) of $\mathbf{A}$, respectively. The expectation operator is denoted by $\mathbb{E}\{\cdot\}$.

%%%%%%%%%%%%%%%%%%%%%%%%%%%%%%%%%%%%%%%%%%%%%%%%%%%%%%%%%%%%%%%%%%%%%%%%%%%%%%%%%%%%%%%%%%%%%%%%%%%%%%%%%%%%%%%%%%%%%%%%%%%%%%%%%%%%%%%%%%%%%%%%%%%%%%%%%%%%%%%%%%%%%%%%%%%%%%%%%%%%%%%%%%%%%%%%%%%%%%%%%%%%%%%%%%%%%%%%%%%%%%%%%%%%%%%%%%%%%%%%%%%%%%%%%%%%%%%%%%%%

\section{System Model}
\label{sec:system_model}

We consider a single-input single-output (SISO) AFDM signal $\mathbf{x} \in \mathbb{C}^N$ composed of $N$ orthogonal chirp subcarriers. When it is sampled at the Nyquist rate, the AFDM signal can be expressed as 

\begin{equation}
    \mathbf{x} = \boldsymbol{\Lambda}_{c_1} \mathbf{F}_{N}^H \boldsymbol{\Lambda}_{c_2} \mathbf{d}, 
    \label{eq:xgen}
\end{equation}
where the vector $\mathbf{d} = [d_0,..,d_{N-1}] \in \mathbb{C}^N$ contains the data randomly taken from a constellation, and $\boldsymbol{\Lambda}_{c_i} = \text{diag}([e^{2 j \pi c_i 0^2},..,e^{2 j \pi c_i (N-1)^2}]) \in \mathbb{C}^{N \times N}$, $i=1,2$, with the post- and pre-chrip parameters  $c_1, c_2 \in \mathbb{R}$, respectively. 

\begin{figure}[b]
\centering{\includegraphics[width=\columnwidth]{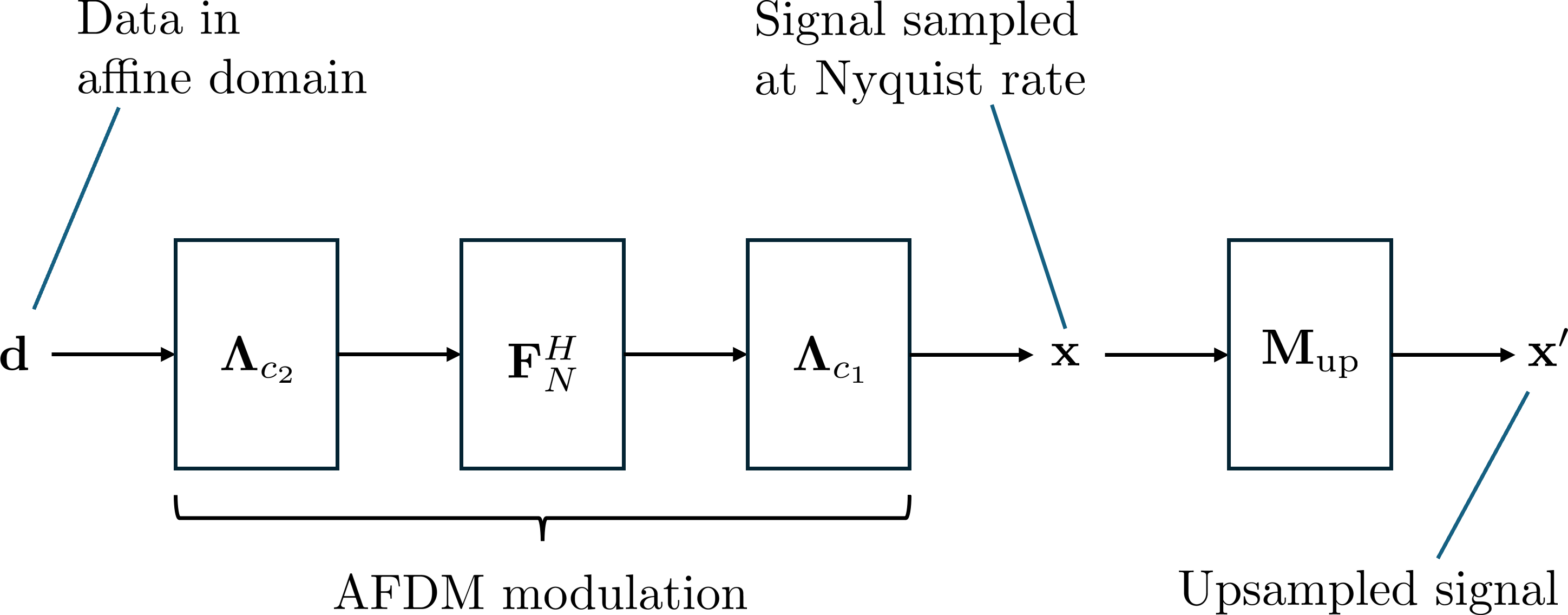}}
\caption{Block diagram of the AFDM transmitter including upsampling.}
\label{fig:system_model}
%\vspace{-2ex}
\end{figure}

A chirp-periodic prefix (CPP) is appended at the beginning of each AFDM symbol to avoid inter-symbol interference introduced by the multipath channel. Let us denote by $N_{\text{CP}}$ the length of the CPP, then for any $n=-N_{\text{CP}},..,-1$, the CPP can be formulated as  

\begin{equation}
    (\mathbf{x})_{n} = (\mathbf{x})_{N+n}e^{-2j\pi c_1(N^2 + 2Nn)},  
    \label{eq:xncp}
\end{equation}
which reduces to the classic cyclic prefix (CP) if $2Nc_1$ is an integer. 
In turn, the oversampled AFDM signal $\mathbf{x}' \in \mathbb{C}^{N'}$ (with $N' > N$) can be expressed from (\ref{eq:xgen}) as follows: 

\begin{equation}
    \mathbf{x}' = \mathbf{M}_{\text{up}} \mathbf{x}, 
    \label{eq:xoversamp}
\end{equation}
where $\mathbf{M}_{\text{up}} \in \mathbb{C}^{N' \times N}$ is the upsampling matrix such that $\frac{N'}{N}$ is the upsampling ratio. Typically, the upsampling can be carried out by appending ${(N' - N)}/{2}$ zero subcarriers at the edges of the signal in the frequency domain, such that $\mathbf{M}_{\text{up}}$ can be obtained as 

\begin{equation}
    \mathbf{M}_{\text{up}} = \mathbf{F}_{N'}^H \mathbf{J} \mathbf{F}_{N}, 
    \label{eq:mup}
\end{equation}
where $\mathbf{J} = [\mathbf{0}^T,\mathbf{I}_N,\mathbf{0}^T]^T$, where $\mathbf{0}$ is a $\frac{N' - N}{2} \times N$ zero matrix. Fig. \ref{fig:system_model} illustrates the AFDM transmitter including upsampling. 

At the receiver side, after downsampling and removing the CPP, the general SISO AFDM input-output relation is given by: 

\begin{equation}
    \mathbf{y} = \underbrace{\sum_{p=1}^{P} 
h_p \, 
\boldsymbol{\Gamma}_{\mathrm{CPP},p} \, 
\boldsymbol{\Pi}^{l_p} \, 
\boldsymbol{\Delta}^{\nu_p}\mathbf{x}}_{\mathbf{H} \mathbf{x}}  + \mathbf{w}, 
    \label{eq:receivedsignaly}
\end{equation}
where $\mathbf{y} \in \mathbb{C}^{N \times 1}$ is the vector of the received signal and $\mathbf{w} \in \mathbb{C}^{N \times 1}$ is the vector of the additive white Gaussian noise such that $\mathbf{w} \sim \mathbb{C}\mathcal{N}(\boldsymbol{0}_N,\sigma^2\mathbf{I}_N)$. 

Next, we define the channel matrix $\mathbf{H} \in \mathbb{C}^{N \times N}$ as the sum over $P$ paths that comprises the path coefficients $h_p \in \mathbb{C}$, while $l_p\in [0,l_{P-1}]$ and $\nu_p \in [-\nu_{\max}, \nu_{\max}]$ denote the delay and Doppler shifts of path $p$. Moreover, we have the matrix $\boldsymbol{\Delta} = \text{diag}([1,e^{2j\pi \frac{1}{N}},..,e^{2j\pi \frac{(N-1)}{N}}]) \in \mathbb{C}^{N \times N}$, and the forward cyclic-shift matrix $\boldsymbol{\Pi}$. 
The matrix $\boldsymbol{\Gamma}_{\mathrm{CPP},p}$ models the effect of the CPP which is given by 

\begin{equation}
\boldsymbol{\Gamma}_{\mathrm{CPP},p}
=
\operatorname{diag}\!\left(
\begin{cases}
e^{-j2\pi c_1 (N^2 - 2N(l_p-n))}, & n<l_p, \\
1, & n\ge l_p.
\end{cases}
\right)
\label{gamacpp}
\end{equation}

Then, the AFDM demodulation matrix $(\boldsymbol{\Lambda}_{c_1} \mathbf{F}_{N}^H \boldsymbol{\Lambda}_{c_2})^{-1}=(\boldsymbol{\Lambda}_{c_1} \mathbf{F}_{N}^H \boldsymbol{\Lambda}_{c_2})^H$ is applied to (\ref{eq:receivedsignaly}), which leads to 

\begin{align}
    \mathbf{r} =& (\boldsymbol{\Lambda}_{c_1} \mathbf{F}_{N}^H \boldsymbol{\Lambda}_{c_2})^{-1} \mathbf{y}  \nonumber \\ 
    =& \underbrace{(\boldsymbol{\Lambda}_{c_1} \mathbf{F}_{N}^H \boldsymbol{\Lambda}_{c_2})^{-1} \mathbf{H} (\boldsymbol{\Lambda}_{c_1} \mathbf{F}_{N}^H \boldsymbol{\Lambda}_{c_2})}_{\mathbf{H}_{\text{eq}}} \mathbf{d} + \underbrace{(\boldsymbol{\Lambda}_{c_1} \mathbf{F}_{N}^H \boldsymbol{\Lambda}_{c_2})^{-1}\mathbf{w}}_{\tilde{\mathbf{w}}}, 
    \label{eq:rafdm}
\end{align}
where the equivalent channel matrix $\mathbf{H}_{\text{eq}}$ is defined for a matter of clarity. Furthermore, we denote the demodulated noise as $\tilde{\mathbf{w}}$, which is white and Gaussian with the same variance as $\mathbf{w}$. 

According to \cite{yin22,bemani23,savaux25_PHYCOM}, the channel $\mathbf{H}_{\text{eq}}$ can be estimated by using pilot subcarriers multiplexed with data in $\mathbf{d}$, and surrounded by a guard intervals to mitigate the interference between pilots and data. Then, the channel can be estimated as presented in \cite{yin22,bemani23,zhou24,zheng25,savaux25_PHYCOM,rou26} and then equalized to obtain 

\begin{equation}
\hat{\mathbf{d}} = \mathbf{G} \mathbf{r} = \mathbf{d} + \mathbf{G}\mathbf{Q}^{-1}\mathbf{w},  
\label{eq:equal}
\end{equation}
from which we extract the recovered data $\hat{\mathbf{d}}$. For instance, if the zero-forcing equalizer is employed, we have $\mathbf{G} = \mathbf{H}_{\text{eq}}^{-1}$. However, we do not further detail the equalization since we mainly focus on the AFDM transmitter in this paper. 

It has been proved in \cite{bemani21,bemani23,rou24} that the full-diversity property of AFDM holds if the chirp parameters $c_1$ and $c_2$ are set as 

\begin{align}
    &\frac{2\nu_{\max}+1}{2N} \leq c_1,\  \quad c_2 << \frac{1}{N}, 
\end{align}
or alternatively $c_2$ should be an irrational number. 

Thus, the choice of the pre-chirp parameter $c_2$ is flexible and can offer a degree of freedom to improve secondary features of the AFDM modulation scheme, such as physical-layer security \cite{Savaux_WCL26,CPAFDM_WCL25,PCS_IWCMC26}, sensing \cite{meng2026unified,OJCOMS_RCAFDM26,cui2026adaptive}, and more \cite{PCDIM_TWC25,CPAFDM_arxiv26}. 
In light of the above, we leverage the flexibility of $c_2$ to jointly reduce the PAPR and the OOBE in the next section.

%%%%%%%%%%%%%%%%%%%%%%%%%%%%%%%%%%%%%%%%%%%%%%%%%%%%%%%%%%%%%%%%%%%%%%%%%%%%%%%%%%%%%%%%%%%%%%%%%%%%%%%%%%%%%%%%%%%%%%%%%%%%%%%%%%%%%%%%%%%%%%%%%%%%%%%%%%%%%%%%%%%%%%%%%%%%%%%%%%%%%%%%%%%%%%%%%%%%%%%%%%%%%%%%%%%%%%%%%%%%%%%%%%%%%%%%%%%%%%%%%%%%%%%%%%%%%%%%%%%%

\section{Pre-Chirp Parameter-Based Joint PAPR and OOBE Reduction}
\label{sec:perf_analysis} 

In this section, we propose a joint PAPR and OOBE reduction based on a Pareto-like optimum obtained in a discrete set. 

%%%%%%%%%%%%%%%%%%%%%%%%%%%%%%%%%%%%%%%%%%%%%%%%%%%%%%%%%%%%%%%%%%%%%%%%%%%%%%%%%%%%%%%%%%%%%%%%%%%%%%%%%%%%%%%%%%%%%%%%%%%%

\subsection{PAPR Definition}

The PAPR, hereby denoted by $\gamma$, is generally defined as 

\begin{equation}
    \gamma =  \frac{\max_n|(\mathbf{x})_n|^2}{\mathbb{E}\{|(\mathbf{x})_n|^2\} }  , 
    \label{eq:PAPR}
\end{equation}
where the denominator is a constant corresponding to the signal power, which is unitary if the signal is normalized as assumed in this paper. 
Because the data elements $d_m$, for $m=0,\ldots,N-1$ are supposed to be independent and identically distributed (iid), $|(\mathbf{x})_n|^2$ are also iid and obey a Chi-square distribution. Then the CCDF of the PAPR can be expressed as 

\begin{align}
    CCDF(\lambda) &= \mathbb{P}\left( \gamma \geq \lambda \right) \nonumber \\
    &= 1-(1-e^{-\lambda})^N,  
    \label{eq:ccdfpapr}
\end{align}
where $\lambda$ is the PAPR threshold. According to \cite{vannee98}, the CCDF of the PAPR of the oversampled signal $\mathbf{x}'$ is more generally given by $CCDF(\lambda) = 1-(1-e^{-\lambda})^{\alpha N}$, where $\alpha$ is empirically set to $\alpha = 2.8$. 

%%%%%%%%%%%%%%%%%%%%%%%%%%%%%%%%%%%%%%%%%%%%%%%%%%%%%%%%%%%%%%%%%%%%%%%%%%%%%%%%%%%%%%%%%%%%%%%%%%%%%%%%%%%%%%%%%%%%%%%%%%%%

\subsection{OOBE Definition}
We define $S_x(f)$ as the "instantaneous" power spectral density of the oversampled CPP-AFDM signal $\mathbf{x}'$ in (\ref{eq:xoversamp}). It is given by the power of the discrete-time Fourier transform (DTFT) $X'(f)$ of $\mathbf{x}'$ as follows: 

\begin{equation}
    S_x(f) =  |X'(f)|^2 , 
    \label{eq:sxf}
\end{equation}
with $X'(f) = \sum_n (\mathbf{x}')_n e^{-2j\pi f t_s n}$, where $t_s$ is the sampling time. Then, the OOBE, denoted by $\beta$, is defined as the integral of (\ref{eq:sxf}) outside the signal bandwidth, which yields: 

\begin{equation}
    \beta = \int_{\Omega_f} S_x(f) \text{d} f, 
    \label{eq:beta}
\end{equation}
where $\Omega_f$ is the frequency range outside the signal bandwidth. 

%%%%%%%%%%%%%%%%%%%%%%%%%%%%%%%%%%%%%%%%%%%%%%%%%%%%%%%%%%%%%%%%%%%%%%%%%%%%%%%%%%%%%%%%%%%%%%%%%%%%%%%%%%%%%%%%%%%%%%%%%%%%

\subsection{Joint PAPR and OOBE Reduction}
\begin{figure}[t]
\centering{\includegraphics[width=0.8\columnwidth]{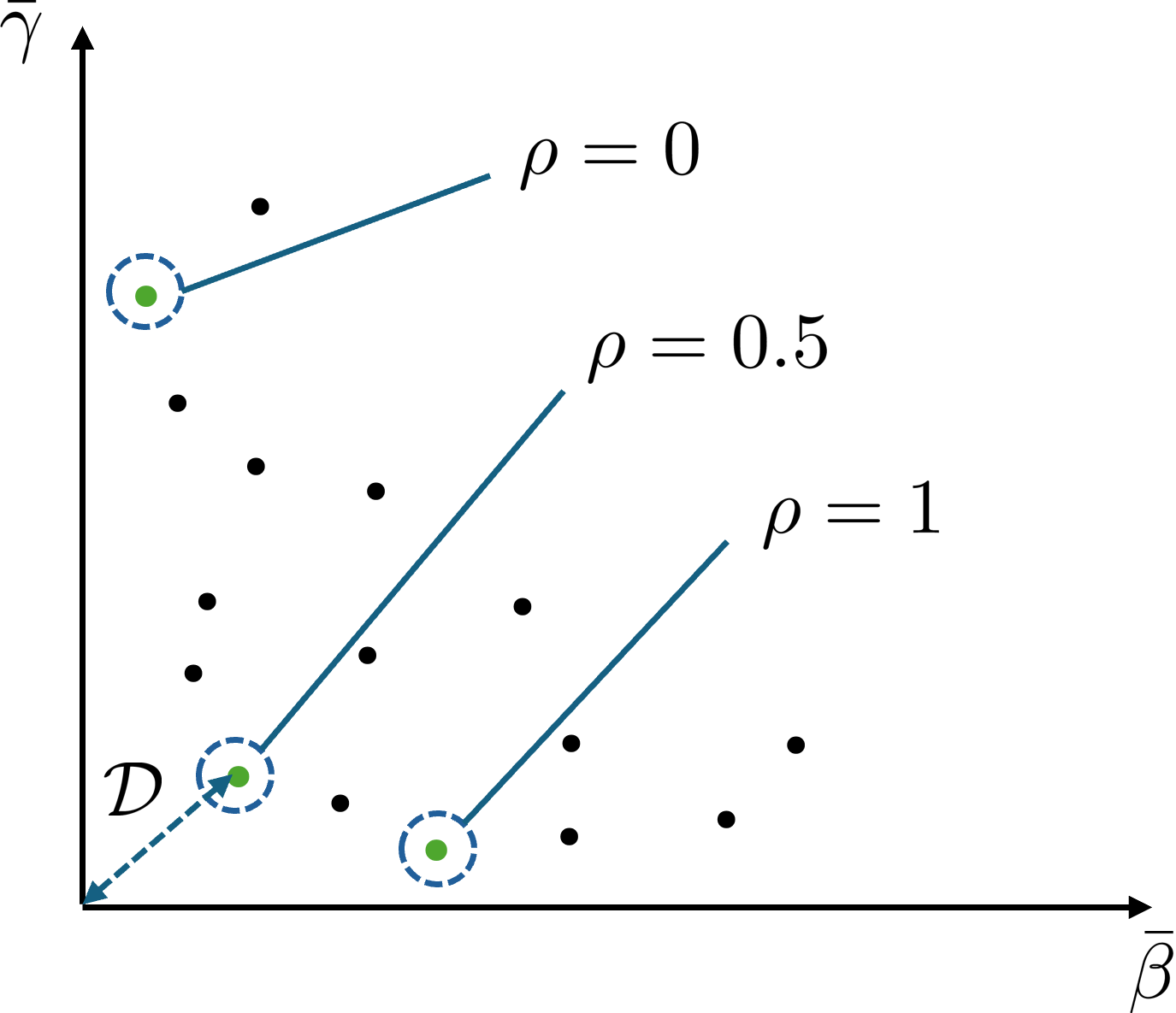}}
\caption{Pareto-like optimum (surrounded dots) for joint PAPR $\bar{\gamma}$ and OOBE $\bar{\beta}$ minimization, with $M_c=16$.}
\label{fig:pareto_optimum}
%\vspace{-2ex}
\end{figure}

In this paper, we consider the pre-chirp selection method (PSM) originally described for PAPR reduction in AFDM in \cite{yuan25}, and similar to the selected mapping in OFDM \cite{bauml96}. The reason for choosing PSM is that it can also be used for the joint reduction of OOBE and PAPR, as described below. Thus, consider the metric $\kappa \in \{\gamma,\beta\}$. The basic principle is to choose the pre-chirp parameter $c_2$ among a set of possible values $\Omega_c = \{c_2^{(0)},c_2^{(1)},..,c_2^{(M_c-1)}\}$ according to the minimization of the cost function $\kappa$ as 

\begin{equation}
    c_2^* = \underset{c_2 \in \Omega_c}{\text{argmin }}  \kappa, 
    \label{eq:minc2}
\end{equation}
where $c_2^*$ is the optimal value, and $M_c$ is the cardinality of $\Omega_c$. It is very difficult to analytically solve (\ref{eq:minc2}) for both $\gamma$ and $\beta$. Instead, we suggest finding the optimal $c_2$ based on a Pareto-like optimum over the discrete set $\Omega_c$. Note that we use the term "Pareto-like optimum" to highlight that we approximate the continuous Pareto front via a discrete set of size $M_c$. We normalize the metrics $\bar{\gamma} = \frac{\gamma}{\mathbb{E}\{\gamma\}}$ and $\bar{\beta} = \frac{\beta}{\mathbb{E}\{\beta\}}$ and define the weighted Euclidian distance $\mathcal{D}$ as 

\begin{equation}
    \mathcal{D} = \sqrt{\rho \bar{\gamma}^2 + (1-\rho)\bar{\beta}^2}, 
    \label{eq:distanceD}
\end{equation}
where $\rho \in [0,1]$ is a weighting factor that can be tuned to support one metric or the other. For instance, $\rho = 1$ corresponds to PAPR only, whereas $\rho = 0$ is OOBE only. When the same importance is given to both metrics, then we have $\rho=0.5$. In any case, the Pareto optimum satisfies: 

\begin{equation}
    c_2^* = \underset{c_2 \in \Omega_c}{\text{argmin }}  \mathcal{D}.  
    \label{eq:minc2pareto}
\end{equation}
The principle of the joint minimization of PAPR and OOBE based on Pareto optimum within our discrete set is illustrated in Fig. \ref{fig:pareto_optimum}, where $M_c=16$. Note that, like SLM in OFDM, the proposed PSM requires side information to transmit the selected $c_2$ value. Alternatively, it can be estimated as in \cite{choi26}, but this is not further addressed in this paper. In the following, the performance of the method is evaluated through simulation. 

%%%%%%%%%%%o /
%%%%%%%%%%%%%%%%%%%%%%%%%%%%%%%%%%%%%%%%%%%%%%%%%%%%%%%%%%%%%%%%%%%%%%%%%%%%%%%%%%%%%%%%%%%%%%%%%%%%%%%%%%%%%%%%%%%%%%%%%%%%%%%%%%%%%%%%%%%%%%%%%%%%%%%%%%%%%%%%%%%%%%%%%%%%%%%%%%%%%%%%%%%%%%%%%%%%%%%%%%%%%%%%%%%%%%%%%%%%%%%%%%%%%%%%%%%%%%%%%%%%%%%%%

\section{Simulation Results}
\label{sec:simu}

In this section, we evaluate the performance of the proposed joint PAPR and OOBE reduction through the CCDF of PAPR and the PSD of the signal. The numerical results have been obtained with MATLAB and averaged over $10^4$ independent Monte Carlo runs. The AFDM signal consists of $N=256$ subcarriers carrying quadratic phase shift keying (QPSK) data elements. The time-domain waveform is generated with chirp parameter $c_1$ such that $2Nc_1 = 4.1$, a CPP of length $N_{\text{CP}} = \frac{N}{8}$. Then, the signal is oversampled by a factor of 4. The pre-chirp parameter $c_2$ is chosen such that $|c_2|<0.01$ in sets $\Omega_c$ of size $M_c \in \{8,32\}$.    

\begin{figure}[tbp]
\centering
  \subfigure[CCDF of PAPR versus $\lambda$ (dB), $M_c = 8$. ]
  {\includegraphics[width=\columnwidth]{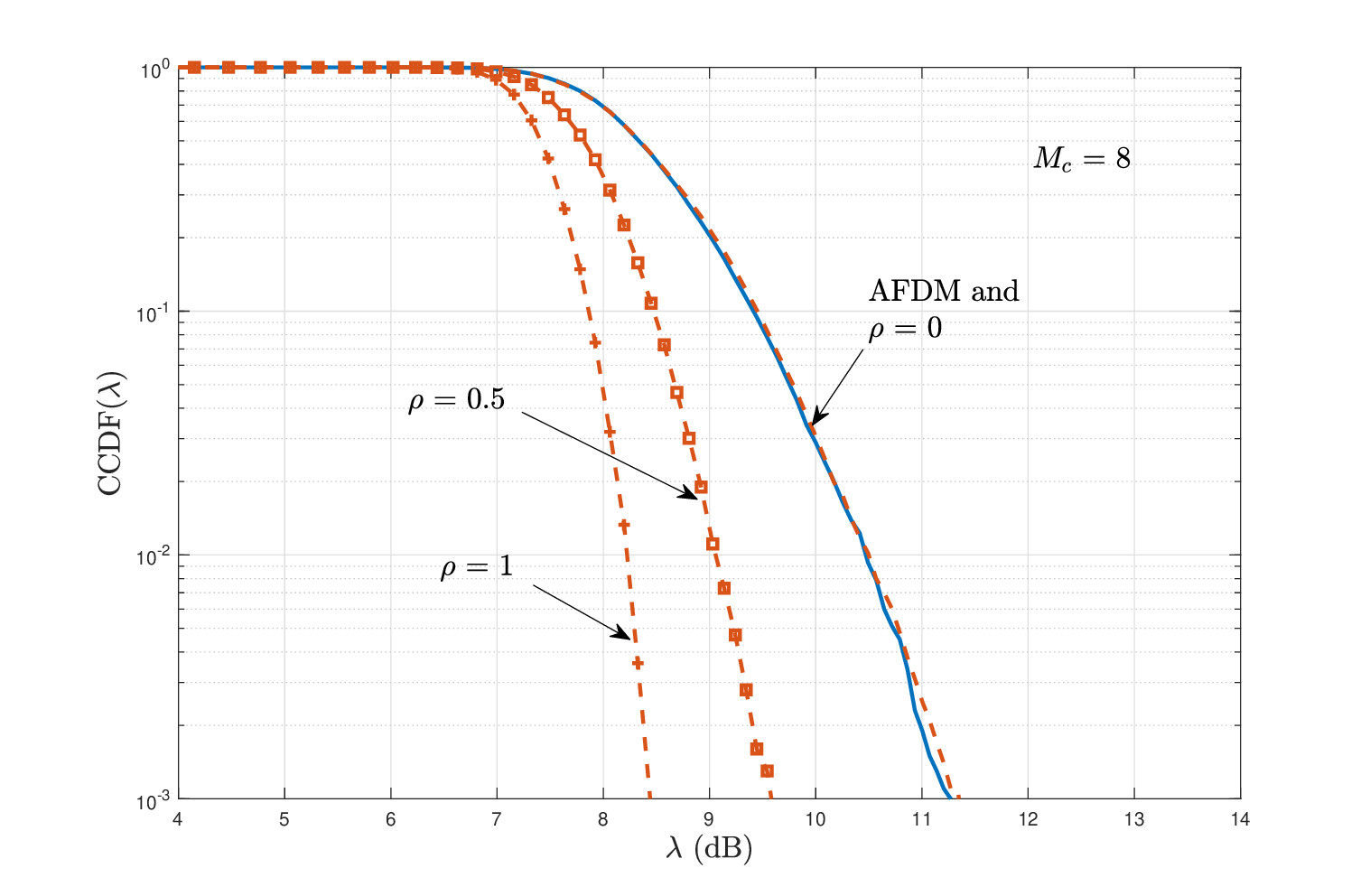}}\quad
  \subfigure[CCDF of PAPR versus $\lambda$ (dB), $M_c = 32$.]
  {\includegraphics[width=\columnwidth]{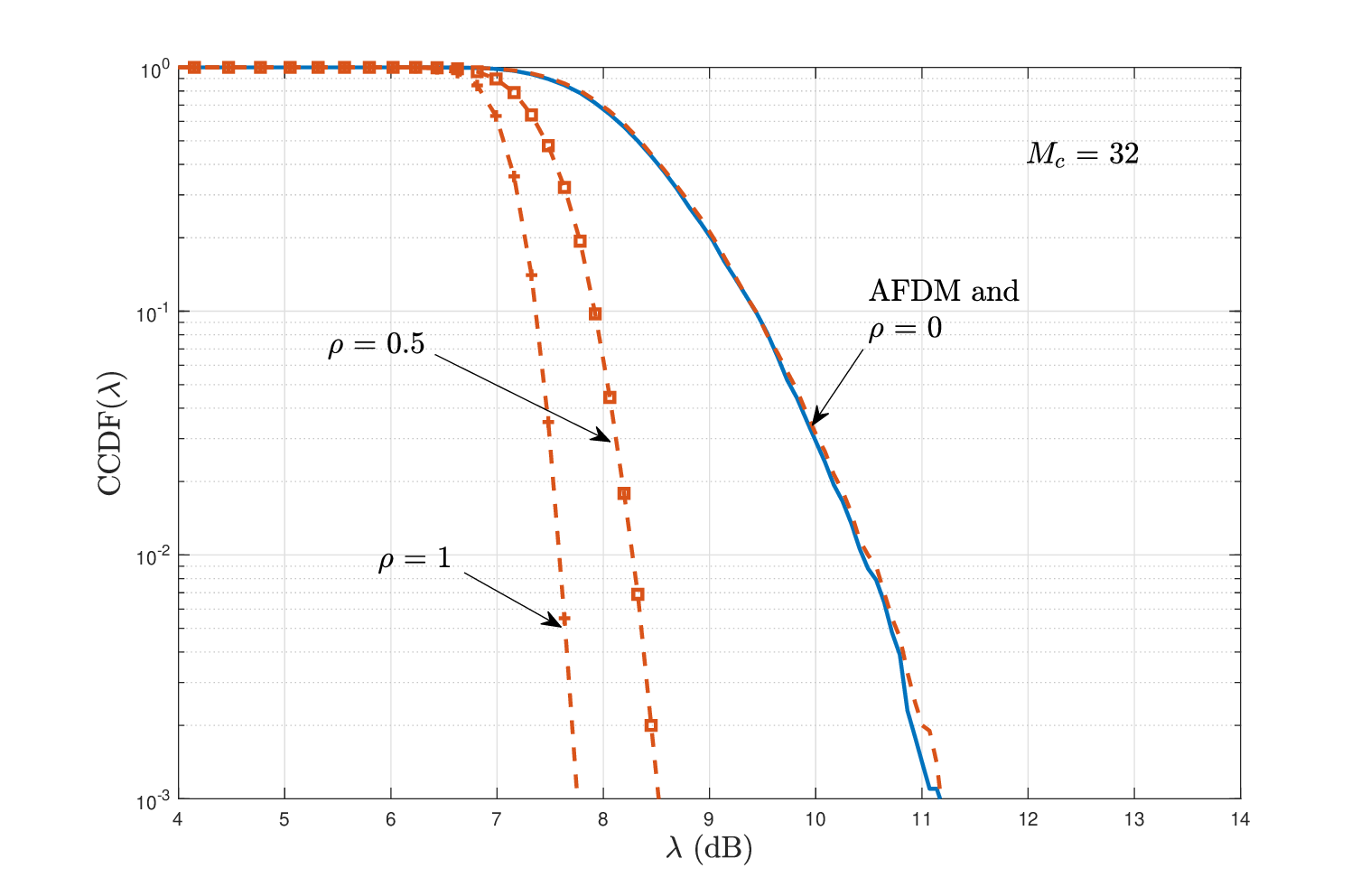}}
\caption{CCDF of PAPR versus $\lambda$ (dB) of AFDM, compared with joint PAPR and OOBE reduction with $\rho =0,0.5,1$, for $M_c = 8$ (a) and $M_c = 32$ (b).}
\label{fig:papr}
%\vspace{-1em}
\end{figure}

Fig. \ref{fig:papr} depicts the CCDF of PAPR versus the threshold $\lambda$ (dB) for the conventional AFDM compared with the proposed optimization method, and two set sizes $M_c=8$ (a) and $M_c=32$ (b). The cases $\rho=0$ and $\rho=1$ correspond to the optimization of the OOBE only and PAPR only, respectively. We also consider $\rho=0.5$, highlighting an equal weighting between OOBE and PAPR for the joint minimization. In Fig. \ref{fig:papr}-(a), it can be observed that $\rho=1$ and $\rho=0.5$ achieve a gain of about 3~dB and 2~dB compared to conventional AFDM (at a given CCDF$=10^{-3}$), respectively, whereas the behavior of $\rho=0$ matches that of AFDM. Thus, $\rho=0.5$ experiences a performance loss of only 1~dB compared to $\rho=1$, which corresponds to the lowest achievable bound since $\rho=1$ is the "PAPR only" minimization. In contrast, no PAPR reduction is achieved when the OOBE only is optimized ($\rho=0$). This behavior is confirmed in Fig. \ref{fig:papr}-(b), where gains of about 3.5~dB and 2.5~dB are achieved by $\rho=1$ and $\rho=0.5$ compared to the conventional AFDM, which shows that the larger the cardinality $M_c$ of $\Omega_c$, the better the PAPR reduction.    

\begin{figure}[tbp]
\centering
  \subfigure[PSD versus normalized frequency, $M_c = 8$. ]
  {\includegraphics[width=\columnwidth]{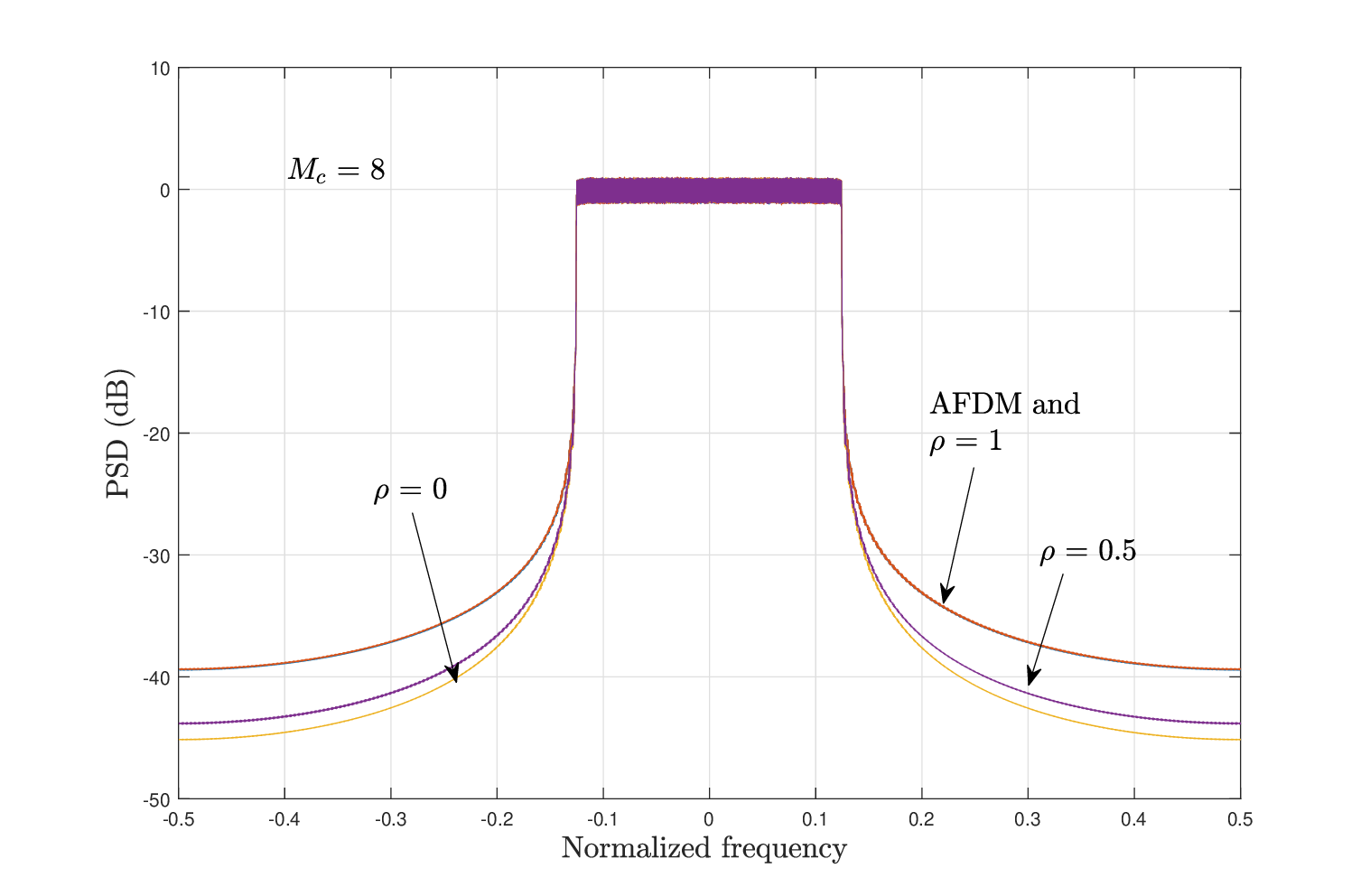}}\quad
  \subfigure[PSD versus normalized frequency, $M_c = 32$.]
  {\includegraphics[width=\columnwidth]{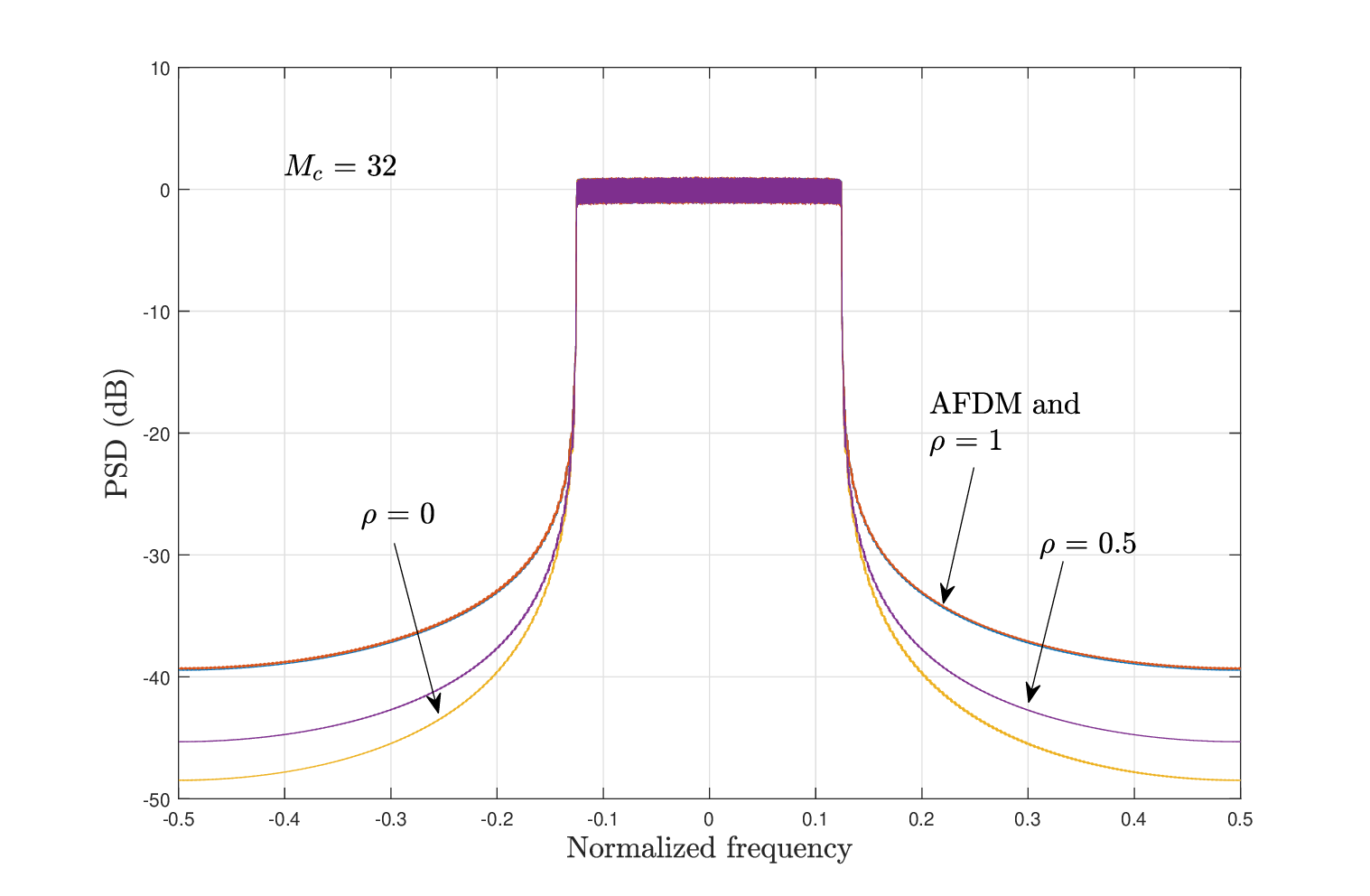}}
\caption{PSD versus normalized frequency, compared with joint PAPR and OOBE reduction with $\rho =0,0.5,1$, for $M_c = 8$ (a) and $M_c = 32$ (b).}
\label{fig:oobe}
%\vspace{-1em}
\end{figure}

Fig. \ref{fig:oobe} shows the PSD versus the normalized frequency of the conventional AFDM compared with the proposed solution, in the same conditions as previously. Again, Fig. \ref{fig:oobe}-(a) and (b) correspond to $M_c=8$ and $M_c=32$, respectively. The PSD of $\rho=1$ matches that of AFDM, as it corresponds to the "PAPR only" scenario, and therefore the OOBE is not impacted. In Fig. \ref{fig:oobe}-(a), we can observe a reduction of the OOBE of about 4.5~dB ($\rho=0.5$) and 6~dB ($\rho=1$, measured at $f=\pm 0.5$) compared to the PSD of the conventional AFDM. In Fig. \ref{fig:oobe}(b), the reduction is up to 6~dB and 9~dB. 

We conclude from Figs. \ref{fig:papr} and \ref{fig:oobe} that: 
\begin{itemize}
    \item Optimizing PAPR only ($\rho=1$) or OOBE only ($\rho=0$) has no advantageous or detrimental impact on the other metric. This is because both metrics are uncorrelated. 
    \item Joint optimization of the PAPR and the OOBE with $\rho=0.5$ (equal weighting) yields substantial gains for both metrics, as they experience only slight performance losses compared to the "one metric only" minimization.   
\end{itemize}

The proposed method then offers a good trade-off between both metric optimizations, which can even be adjusted through $\rho$ according to the importance that the system designer attaches to PAPR or OOBE. 

%%%%%%%%%%%%%%%%%%%%%%%%%%%%%%%%%%%%%%%%%%%%%%%%%%%%%%%%%%%%%%%%%%%%%%%%%%%%%%%%%%%%%%%%%%%%%%%%%%%%%%%%%%%%%%%%%%%%%%%%%%%%%%%%%%%%%%%%%%%%%%%%%%%%%%%%%%%%%%%%%%%%%%%%%%%%%%%%%%%%%%%%%%%%%%%%%%%%%%%%%%%%%%%%%%%%%%%%%%%%%%%%%%%%%%%%%%%%%%%%%%%%%%%%%%%%%%%%%%%%

\section{Conclusion}
\label{sec:conclusion} 

In this paper, we propose a joint PAPR and OOBE reduction method based on selecting the chirp parameter $c_2$. To this end, we first introduce a distance defined as the weighted sum of the PAPR and the OOBE. Then, $c_2$ is selected by minimizing this new metric, which is equivalent to finding the Pareto optimum within a discrete set of possible solutions in a two-dimensional problem. The performance of the proposed solution is evaluated through simulation results, which show that both PAPR and OOBE can be reduced, leading to performance close to that of methods specifically optimized for a single objective. This paper paves the way for further studies where other objective functions could be included in the joint optimization problem.

\bibliographystyle{IEEEtran}
%\bibliography{mabiblio}
\bibliography{rev_biblio}

\end{document}